# Analogs of wave function reduction, quantum entanglement and EPR experiment in classical physics of spacetimes with time machines

I. A. Ovid'ko

Department of Mathematics and Mechanics, St. Petersburg State University,

St. Petersburg 198504, Russia

Email:  ovidko@nano.ipme.ru

It is theoretically revealed that, in classical physics of spacetimes with wormholes, there are analogs of wave function reduction events, quantum entanglement and Einstein-Podolsky-Rosen (EPR) experiment. Within the suggested approach, wormholes are specified by a typical microscopic radius of their mouths, and this causes the size effect in operation of wormhole-based time machines (closed timelike curves; CTCs). For geometric reasons, classical solid balls in a spacetime with a wormhole are divided into the two categories: small and large balls whose traverse through wormholes is permitted and forbidden, respectively. Evolutions of small balls on CTCs can be self-inconsistent (or, in other terms, inconsistent with conventional causality), in which case there is an uncertainty in their behaviors. In contrast, evolutions of large solid balls are always unambiguous. In the situation where small balls can be absorbed by large balls, uncertain behaviors of small balls transform into unambiguous evolutions of large balls in the logical way analogous to that of a quantum measurement - wave function reduction - event. Also, within the suggested approach operating with classical balls in spacetimes with wormholes, analogs of quantum entanglement and EPR experiment are defined and theoretically described.

## I. INTRODUCTION

Operation of time machines in spacetimes with wormholes [1-3] generates causality paradoxes which are of utmost interest from a fundamental viewpoint; see, e.g., [3-10]. More precisely, wormhole-based time machines or, in other terms, CTCs in principle allow one to travel to the past and thereby violate conventional causality through change of the past. For a long time, the idea on such causality paradoxes has been effectively exploited in science fiction. The most famous example is the "grandfather paradox" (a person traveled back in time and killed his grandfather before the latter met the traveler grandmother); see, e.g., discussions in papers [8,9].

An elegant approach to avoid the causality violation generated by CTCs in classical physics was suggested by Friedman et al [3]. This approach is based on the principle of self-consistency which admits the only self-consistent evolutions on CTCs in the sense that these evolutions change the past in the way keeping them unambiguous [3]. Other evolutions - self-inconsistent evolutions which by definition do not satisfy the principle of self-consistency and thereby violate the conventional causality - are not allowed to be realized in nature [3]. The principle of self-consistency was successfully exploited in selection and theoretical description of self-consistent evolutions of perfectly elastic solid balls ("billiard balls" serving as simple models of classical particles) and other classical systems in spacetimes with CTCs [4,6]. It is interesting to note that the principle of self-consistency [3] is equivalent to "banana peel mechanism" preventing the grandfather paradox in science fiction. In its terms, if a time machine operates, there always exists a strategically placed banana peel on which the prospective murderer slips as he pulls the trigger, thus spoiling his arm [9].

Deutsch [5] considered physical effects of CTCs on evolutions of quantum-mechanical systems. Within the approach [5], self-inconsistent evolutions of quantum systems in spacetimes with CTCs automatically do not come into play in the Everett multiworld interpretation of quantum mechanics. More precisely, the pairs of seemingly inconsistent events are realized in "different" universes, in which case the events are interpreted as consistent [5] (see also a discussion in papers [9,10]). In this context, following Deutsch [5], it is possible to experimentally distinguish the Everett multiworld interpretation from other interpretations of quantum mechanics in spacetimes with CTCs. Also, it is interesting to note that the Deutsch approach [5] is similar to multiworld interpretations of seemingly inconsistent classical events (e.g., birth of a person and "this person kills his young grandfather" event) at time loops in science fiction; see, e.g., [9]. In the classical or science-fiction multiworld interpretation, the pairs of seemingly inconsistent events are not in contradiction, because they occur in "different" universes.

Thus, the traditional physical concepts - the principle of self-consistency [3] and the Deutsch approach [5] - dealing with causality paradoxes in spacetimes with CTCs are similar to those - "banana peel mechanism" and "classical multiworld", respectively - applied to solution of the grandfather paradox and its analogs in science fiction. Recently, Ovid'ko [11] has suggested an alternative approach taking into account the size effect (finite sizes of wormhole mouths) in classical physics of spacetimes with CTCs and having a qualitative similarity to quantum mechanics as the theory of microparticles interacting with classical macrosystems. The alternative approach is concerned with small and large classical balls that traverse and cannot traverse through wormholes, respectively. The focus is placed on self–inconsistent evolutions of small classical particles traversing through CTCs and interacting with large classical particles whose evolutions are unambiguous [11]. The previous paper [11] briefly presented the basic statements of the alternative approach. The main aims of this research are to give its extended presentation, discuss its logical similarity to quantum mechanics (outside Everett interpretation) and theoretically describe analogs of wave function reduction events, quantum entanglement and EPR experiment in classical physics of spacetimes with CTCs.

# II. STANDARD PRINCIPLE OF SELF-CONSISTENCY IN SPACETIME WITH A WORMHOLE-BASED TIME MACHINE: SELF-CONSISTENT AND SELF-INCONSISTENT EVOLUTIONS OF CLASSICAL SOLID BALLS

In this section, following the approach [3], we will illustrate the conventional principle of self-consistency in the simplest case of a perfectly elastic solid ball moving in a spacetime with a CTC associated with a static wormhole (Fig. 1). The wormhole has two mouths A and B, spherical holes shown as cross-hatched circles in Fig. 1. They are characterized by the same radius $R$ and the distance $D$ between them in a three-dimensional space. In addition, the mouths are connected by a short handle with the negligibly small length $l \ll D$. For the aims of this paper focused on causality paradoxes, it is taken that $l = 0$. The spacetime under examination is a flat spacetime everywhere except for the wormhole mouths and their vicinities.

Let us consider perfectly elastic solid balls each having radius $r < R$ and moving in the spacetime with the wormhole (Figs. 1 and 2). Following the theory of spacetimes with wormholes [3], in the discussed case, there are CTCs which involve traverse through the wormhole. The latter means that, when a solid ball of radius $r < R$ enters the wormhole mouth A, the ball appears from the mouth B in the past (Figs. 1 and 2).

In general, the existence of CTCs can generate violations of conventional causality. This aspect is well illustrated with the notions of self-consistent and self-inconsistent evolutions of solid balls traversing through the wormhole [3]. Following the approach [3], let us consider a typical example of self-consistent evolution of a solid ball hereinafter called the ball 1 (Fig. 1a). First, the ball 1 moves along the trajectory $\alpha$, and undergoes self-collision - collision with "older itself" - at point O (Fig. 1a). Then (in proper time of the ball), the ball 1 moves along the trajectory $\alpha'$, reaches the wormhole mouth A at the end of the trajectory $\alpha'$, enters the mouth A and appears from the wormhole mouth B in the past (Fig. 1a). Then, the ball 1 moves along the trajectory $\beta$ and collides with "younger itself" at point O (Fig. 1a). After the collision, the ball moves along the trajectory $\beta'$ (Fig. 1). If details of the self-consistent evolution along a CTC are not significant, one can describe it as that occurring within a black box with the trajectory $\alpha$ coming into the box from the past and the trajectory $\beta'$ moving from the box towards the future (Fig. 1b).

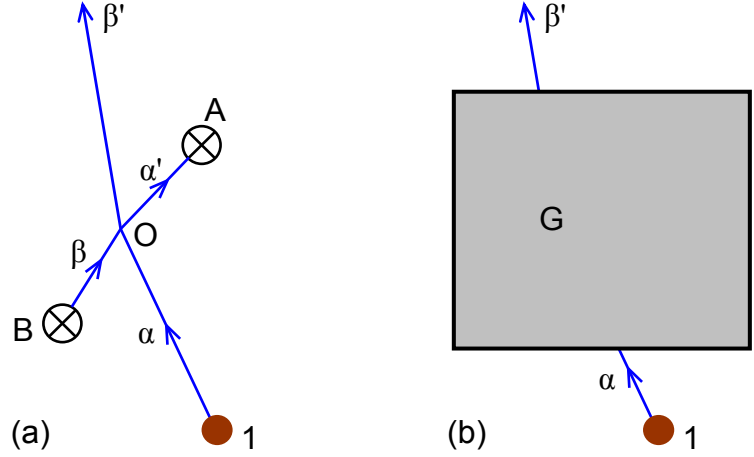

Fig. 1. (Color online) Self-consistent evolution of a classical solid ball in spacetime with a static wormhole: a typical example. (a) The solid ball (full circle) 1 moves along the trajectory $\alpha$, collides with older itself at point O, moves along the trajectory $\alpha'$ and reaches the wormhole mouth A (cross-hatched circle) at the end of the trajectory $\alpha'$. The ball 1 enters the mouth A, then appears from the wormhole mouth B (cross-hatched circle) in the past, moves along the trajectory $\beta$, collides with younger itself at point O, and then moves along the trajectory $\beta'$. The ball trajectories $\alpha$ and $\beta$ meet at point O where the ball collides with itself. As a result of this self-collision, the ball moving along trajectories $\alpha$ and $\beta$ drives itself to move along trajectories $\alpha'$ and $\beta'$, respectively, in which case the trajectory $\alpha'$ ends at the wormhole mouth A, and the ball 1 moves along the trajectory $\beta'$ towards the future. (b) If details of the self-consistent evolution on the wormhole-based CTC are not interesting, one can consider it as that occurring within a black box (shown as a grey box) with the trajectory $\alpha$ coming into the box from the past and the trajectory $\beta'$ moving from the box towards the future.

Note that, for any external observer, the ball 1 has the conventionally defined past, present and future at the trajectory $\alpha O \beta'$. In this context, self-consistent evolutions of solid balls do not violate conventional causality and, in accord with the conventional principle of self-consistency, are allowed to be realized in spacetimes with CTCs [3].

Now let us discuss, following the approach [3], a typical example of self-inconsistent evolution of a solid ball in the spacetime with the wormhole (Fig. 2a). Let a small solid ball (hereinafter denoted as the ball 2) move in the spacetime with the wormhole as follows. The ball 2 moves along the trajectory $\alpha$, enters the wormhole mouth A at the end of this trajectory and appears from the mouth B in the past (Fig. 2a). Then, the ball 2 moves along the trajectory $\beta$, and the ball trajectories $\alpha$ and $\beta$ meet at point O where the ball 2 collides with itself (Fig. 2a). As a result of this self-collision, the trajectories $\alpha$ and $\beta$ of the ball 2 transform into the trajectories $\delta$ and $\gamma$, respectively, which do not end at the wormhole mouth A (Fig. 2a). If it is so, the ball 2 does not move along the wormhole handle AB, does not appear at the mouth B, does not collide with itself at point O; and, as a corollary, the ball 2 moves along the trajectory $\alpha$, enters the mouth A at the end of this trajectory, and so on (Fig. 2a). To summarize, the ball enters the wormhole mouth A, if and only if it does not enter the wormhole mouth A. Thus, the self-inconsistent evolution (Fig. 2a) produces the paradox which is called the Polchinskii paradox [3] and serves as a direct analog of the grandfather paradox in science fiction.

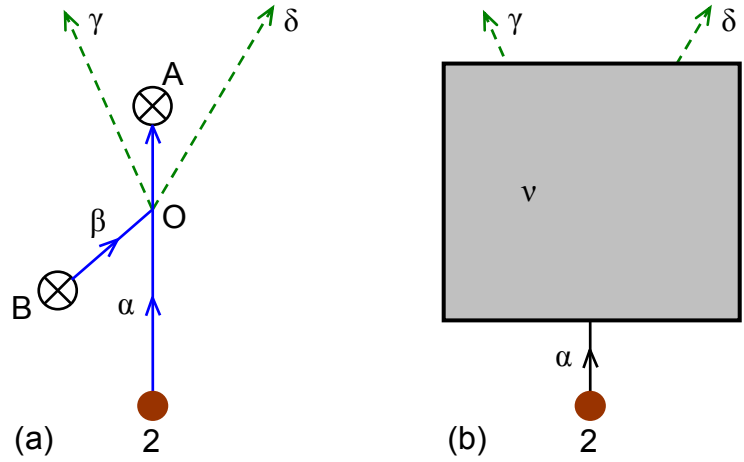

Fig. 2. (Color online) Self-inconsistent evolution of a classical solid ball in spacetime with a wormhole: a typical example. (a) The solid ball (full circle) 2 moves along the trajectory $\alpha$ and enters the wormhole mouth A (cross-hatched circle) at the end of this trajectory. Then the ball appears from the wormhole mouth B (cross-hatched circle) in the past, moves along the trajectory $\beta$. The ball trajectories $\alpha$ and $\beta$ meet at point O where the ball collides with itself. As a result of this self-collision, the ball moving along trajectories $\alpha$ and $\beta$ drives itself to move along trajectories $\delta$ and $\gamma$, respectively, in which case both the trajectories $\gamma$ and $\delta$ do not end at the wormhole mouth A. If it is so, the ball does not move along the wormhole handle AB, does not appear at the mouth B, does not collide with itself at point O and, as a corollary, moves along the trajectory $\alpha$, enters the mouth A at the end of this trajectory, and so on. That is, the ball enters the wormhole mouth A, if and only if it does not enter the wormhole mouth A. (b) If details of the self-inconsistent evolution on the wormhole-based CTC are not interesting, one can consider it as that occurring within a black box with the unambiguous trajectory $\alpha$ coming into the box from the past and the two ambiguous trajectories $\gamma$ and $\delta$ moving from the box towards the future (for details, see text).

If details of the self-inconsistent evolution along a CTC are not significant, one can represent it as that occurring within a black box with the trajectory $\alpha$ coming into the box from the past and both the trajectories $\gamma$ and $\delta$ moving towards future (Fig. 2b), in which case realizations of the trajectories $\gamma$ and $\delta$ are uncertain due to the Polchinskii paradox. Hereinafter we will denote the trajectories $\gamma$ and $\delta$ as "ambiguous trajectories" in order to distinguish them from classical unambiguous trajectories with well defined past, present and future. Also, thereinafter, the same term "ambiguous trajectory" will designate any uncertain trajectory resulted from a self-inconsistent evolution of a classical ball in a spacetime with CTCs.

The ball 2 at ambiguous trajectories of its self-inconsistent evolution does not have any conventionally defined past, present and future. Therefore, self-inconsistent evolutions of classical balls (Fig. 2a) violate the conventional causality. With this aspect, the standard principle of self-consistency [3] states that self-inconsistent evolutions are impossible in principle.

## III. MODIFIED PRINCIPLE OF SELF-CONSISTENCY AND SIZE EFFECT ON SELF-INCONSISTENT EVOLUTIONS. ANALOGS OF WAVE FUNCTION REDUCTION EVENTS IN PHYSICS OF CLASSICAL SOLID BALLS MOVING IN SPACETIME WITH A STATIC WORMHOLE

The standard principle of self-consistency [3] deals with perfectly elastic solid balls traversing a wormhole in a classical spacetime. Recently, a modified principle of self-consistency has been suggested in a more complication situation [11]. The modified principle [11] operates with solid balls of two types - small and large balls having their radii smaller and larger than the wormhole mouth radius, respectively - and involves in consideration specific inelastic collisions between small and large balls.

Following [11], small and large classical balls in a spacetime with a wormhole are defined as follows. All the small solid balls have the same radius $r$ smaller than the wormhole mouth radius $R$. Since $r < R$, these small balls can in principle traverse through the wormhole with radius $R$, and their evolutions can be self-inconsistent.

All the large solid balls are characterized by the same radius $R_L$ larger than the radii of the wormhole mouths and small balls: $R_L>R>r$. As a corollary, in contrast to small balls, large solid balls cannot traverse through the wormhole with the mouth radius $R < R_L$. In these circumstances, large balls can not have ambiguous evolutions; they always have unambiguous evolutions with well defined past, present and future in the spacetime with the wormhole.

The inequalities $R_L>R>r$ describe the size effect in the spacetime with the wormhole. This effect dramatically influences behaviors of solid balls. (It is rather logical, taking into account the significant role of size effects in many other physical systems, say, nanostructured solids; see, e.g., [12-17]). In particular, with the size effect, small classical balls can show uncertain behavior inconsistent with conventional causality. On the first glance, self-inconsistent evolutions of small classical balls are in conflict with our everyday experience. However, in modern physics, there is a remarkable example where uncertain behavior is typical and comes into play due to the size effect. It is the case of quantum microparticles showing uncertainty in their behavior (see, e.g., [18,19]), qualitatively similar to uncertainty in the behavior exhibited by small classical balls in the spacetime with CTCs. Uncertain behaviors of both small classical balls, after their traverse through wormholes, is contrasted to conventional behavior of large classical balls. This contrast resembles that between quantum microparticles and macrosystems in quantum mechanics. In fact, quantum mechanics in its basis is formulated as the theory of microparticles interacting with classical macrosystems. In the context discussed, in our following examination of uncertain behavior of small classical balls, a special attention will be devoted to the interactions between small and large classical balls in the spacetime with CTCs.

As the starting point, let us consider specific collisions between small and large balls, the namely collisions at which small balls are absorbed by large balls. After a collision, a small ball becomes a part of a large ball. That is, the specific collisions provide a link between categories of classical solid balls showing dramatically different individual behaviors in the spacetime with CTCs.

In spirit of our everyday experience, large classical balls are macroscopic objects exhibiting unambiguous behavior. Therefore, after a collision of a large ball and a small ball, the trajectory of the large ball should be unambiguous. However, with ambiguous character of the trajectories of the small ball before the collision event, the large ball trajectory has two potential versions of its realization (Fig. 3).

As to details, first, let us consider a probe situation where the large ball L has its unambiguous trajectory κ in the absence of other balls (Fig. 3a). Second, consider another situation with the large ball L and the small ball S moving in the spacetime with a wormhole. The ball S moves along ambiguous trajectories $\gamma$ and $\delta$ which intersect the trajectory κ in two points $O_\gamma$ and $O_\delta$, respectively (Fig. 3b and c). The ball L moves along the trajectory κ until its meeting with the ball S (Fig. 3b and c). With ambiguous character of the trajectories $\gamma$ and $\delta$, one can not unambiguously predict the point where the balls L and S meet. It may be either point $O_\delta$ (Fig. 3a) or $O_\gamma$ (Fig. 3b). If the balls L and S meet at point $O_\delta$, the large ball absorbs the small ball at this point, and the resultant large ball moves along the trajectory $κ_\delta$ (Fig. 3a). If the balls *L* and *S* meet at point $O_\gamma$, the ball *L* absorbs the ball *S* at this point, and the resultant large ball moves along the trajectory $κ_\gamma$ (Fig. 3b). With the ambiguous character of the trajectories $\gamma$ and $\delta$, the system chooses one of the evolution variants as the real evolution. However, which variant will be chosen is unknown in advance.

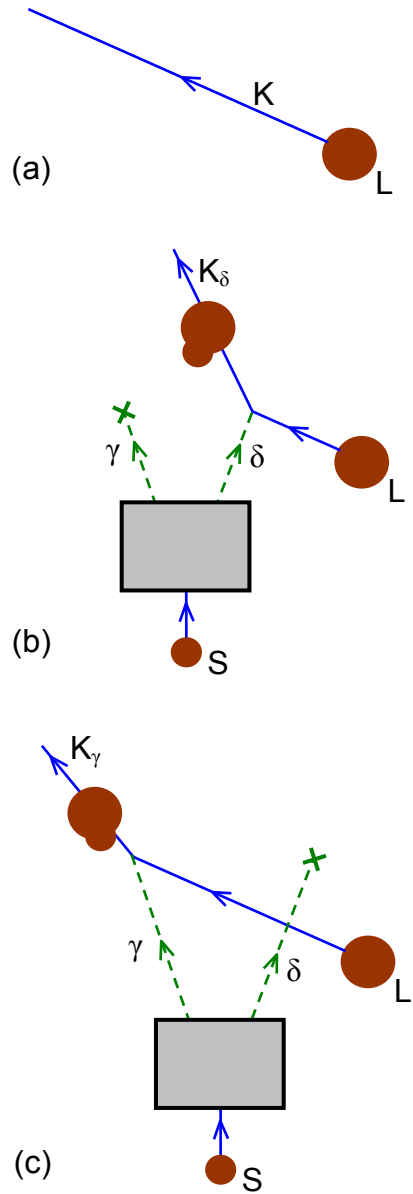

Fig. 3. (Color online) Specific collision of large ball L and small solid ball S in spacetime with a wormhole serves as an analog of quantum measurement event in quantum mechanics (for details, see text). (a) A probe situation where the large ball L has its unambiguous trajectory κ in the absence of other balls. (b) and (c) A collision of a large ball L having unambiguous trajectory κ with a small solid ball S having two ambiguous trajectories $\gamma$ and $\delta$ in spacetime with a wormhole. The collision is accompanied by absorption of ball S by ball L. With ambiguous character of the trajectories $\gamma$ and $\delta$, two versions of the absorption event are possible. (b) If the balls L and S meet at point $O_\delta$, the large ball absorbs the small ball at this point, and the resultant large ball moves along the trajectory $\kappa_\delta$. Cross shows the end of the trajectory $\gamma$ that does not result in the absorption event. (c) If the balls *L* and *S* meet at point $O_\gamma$, the ball *L* absorbs the ball *S* at this point, and the resultant large ball moves along the trajectory $\kappa_\gamma$. Cross shows the end of the trajectory $\delta$ that does not result in the absorption event.

The discussed logical scheme for description of collisions of large and small balls, involving the absorption process (Fig. 3b and c), is analogous to that exploited in description of quantum measurement - wave function reduction – events in quantum mechanics. In doing so, small and large classical balls in a spacetime with CTCs play the roles as analogs of microparticles and macroscopic objects in quantum mechanics, respectively. That is, in at least a rough approximation, physics of classical balls in the spacetime with CTCs in its basis can be interpreted in terms of quantum mechanics and vice versa. In order to illustrate the interpretation in question, in next section, within physics of classical balls in the spacetime with CTCs, we will consider analogs of two remarkable phenomena reflecting the intrinsic nature of quantum mechanics, the namely quantum entanglement and EPR experiment.

**IV. ANALOGS OF QUANTUM ENTANGLEMENT AND EINSTEIN-PODOLSKY-ROSEN EXPERIMENT IN PHYSICS OF CLASSICAL SOLID BALLS MOVING IN SPACETIME WITH A STATIC WORMHOLE**

Let us consider behavior of small classical balls interacting through elastic collisions in the spacetime with a static wormhole. In a probe situation, a small solid ball (hereinafter denoted as the ball $S_1$) has its unambiguous trajectory $\sigma$ in the absence of other balls (Fig. 4a). Now let us examine another situation where the small ball $S_1$ and another small ball $S_2$ move in the spacetime with the wormhole. The ball $S_2$, after its traverse through the wormhole, moves along ambiguous trajectories $\gamma$ and $\delta$. The ambiguous trajectory $\gamma$ of the small ball $S_2$ meets a conventional trajectory $\sigma$ of the ball $S_1$ at some point P of the spacetime (Fig. 4b). Due to uncertainty in realization of the ambiguous trajectory $\gamma$ of the ball $S_2$, its collision with another solid ball at point P is uncertain, too. With this uncertainty, the unambiguous trajectory $\sigma$ of the ball $S_1$ ends at point P where its evolution has become uncertain (Fig. 4b). In terms of trajectories, the trajectory $\sigma$ of the ball $S_1$ at point P transforms into the two ambiguous trajectories $\sigma_1$ and $\sigma_2$ (Fig. 4b). (The trajectory $\sigma_1$ corresponds to the no-collision evolution variant, whereas the trajectory $\sigma_2$ to the variant with the collision between the balls.) Thus, an uncertain collision of a small ball having a classical unambiguous trajectory and a small ball having ambiguous trajectories in the spacetime with CTCs leads to transformation of the unambiguous trajectory of the former ball into two its ambiguous trajectories (Fig. 4 b).

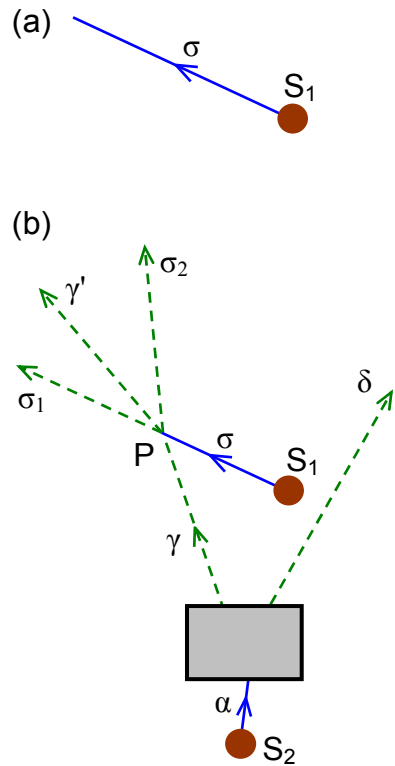

Fig. 4. (Color online) A collision of two small solid balls in spacetime with a wormhole serves as an analog of quantum entanglement in quantum mechanics (for details, see text). (a) A probe situation where small solid ball $S_1$ has its unambiguous trajectory $\sigma$ in the absence of other balls. (b) A collision of two small solid balls $S_1$ and $S_2$ moving in the spacetime with wormhole. The ball $S_2$ traverses through the wormhole, and it is schematically shown as the traverse through a grey box. After the traverse, the ball $S_2$ moves along ambiguous trajectories $\gamma$ and $\delta$. The ambiguous trajectory $\gamma$ of the small ball $S_2$ meets a conventional trajectory $\sigma$ of the ball $S_1$ at some point P of the spacetime. Due to uncertainty in realization of the ambiguous trajectory $\gamma$ of the ball $S_2$, its collision with another solid ball at point P is uncertain, too. With this uncertainty, the trajectory $\sigma$ of the ball $S_1$ at point P transforms into the two ambiguous trajectories $\sigma_1$ and $\sigma_2$.

In interpretation of small classical balls in the spacetime with CTCs as analogs of quantum microparticles, the discussed uncertain collisions between small classical balls (Fig. 4 b) serve as analogs of quantum entanglement. That is, after the uncertain collision (Fig. 4 b), behaviors of the small balls become correlated in the manner analogous to that in the case of entangled microparticles in quantum mechanics. In order to illustrate the analogy in question, let us consider an imaginary experiment with four classical balls, whose schematic illustration is presented in Fig 5. This experiment with classical solid balls in the spacetime with a wormhole (Fig. 5) is similar to EPR experiment in quantum mechanics.

As to details, we consider four classical balls moving in the spacetime with a static wormhole: the large balls $L_1$ and $L_2$ with the unambiguous trajectories ρ and ν, respectively; the small ball $S_1$ with initially unambiguous trajectory σ; and the small ball $S_2$ that moves along the ambiguous trajectories γ and δ (Fig. 5). The initially unambiguous trajectory σ of the small ball $S_1$ meets the ambiguous trajectory γ of the small ball $S_2$ at point P where the trajectory σ transforms into the two ambiguous trajectories $σ_1$ and $σ_2$ (Fig. 5). After that, the system under examination has the following two variants of its further evolution. According to the first variant, the small ball $S_1$ and the large ball $L_1$ collide at point $P_1$ where the ball $S_1$ is absorbed by the ball $L_1$, and then the ball $L_1$ moves along the trajectory $ρ_1$ (Fig. 5a). In addition, the ball $S_2$ (whose ambiguous trajectory γ at point F transforms into unambiguous one) is absorbed by the ball $L_2$ at point G. In the second evolution variant, the small ball $S_1$ and the large ball $L_1$ collide at point $P_2$ where the ball $S_1$ is absorbed by the ball $L_1$, and then the ball $L_1$ moves along the trajectory $ρ_2$ (Fig. 5b). Besides, the ball $S_2$ moves along the trajectory γ' (that becomes unambiguous at time moment at which the ball $S_1$ is absorbed by the ball $L_1$), and it is not absorbed by the ball $L_2$ at point G.

Note that the traverse of the large ball $L_2$ through point G at the trajectory ν may lie outside the light cone of the traverse of the ball $L_1$ through point $P_1$ in the first evolution variant (Fig. 5a). That is, these traverse events may be spacelike separated ones. Also, the ball $L_2$ traverse through point G may lie outside the light cone of the ball $L_1$ traverse through point $P_2$ in the second evolution variant (Fig. 5b). If it is so, the selection of the first or second evolution variant for the ball $L_1$ and selection of a change or constancy of the ball $L_2$ movement direction at point G (Fig. 5) are spacelike separated events. At the same time, the absorption of the ball $S_1$ at point $P_1$ occurs, if the ball $L_2$ changes its movement direction at point G, and vice versa (Fig. 5). Also, the absorption of the ball $S_1$ at point $P_2$ occurs, if the ball $L_2$ does not change its movement direction at point G, and vice versa (Fig. 5). That is, the spacelike separated events occur in a correlated manner (Fig. 5). In the context discussed, the evolution of classical balls with its variants presented in Figure 5 in the spacetime with CTCs serves as an analog of EPR experiment in quantum mechanics.

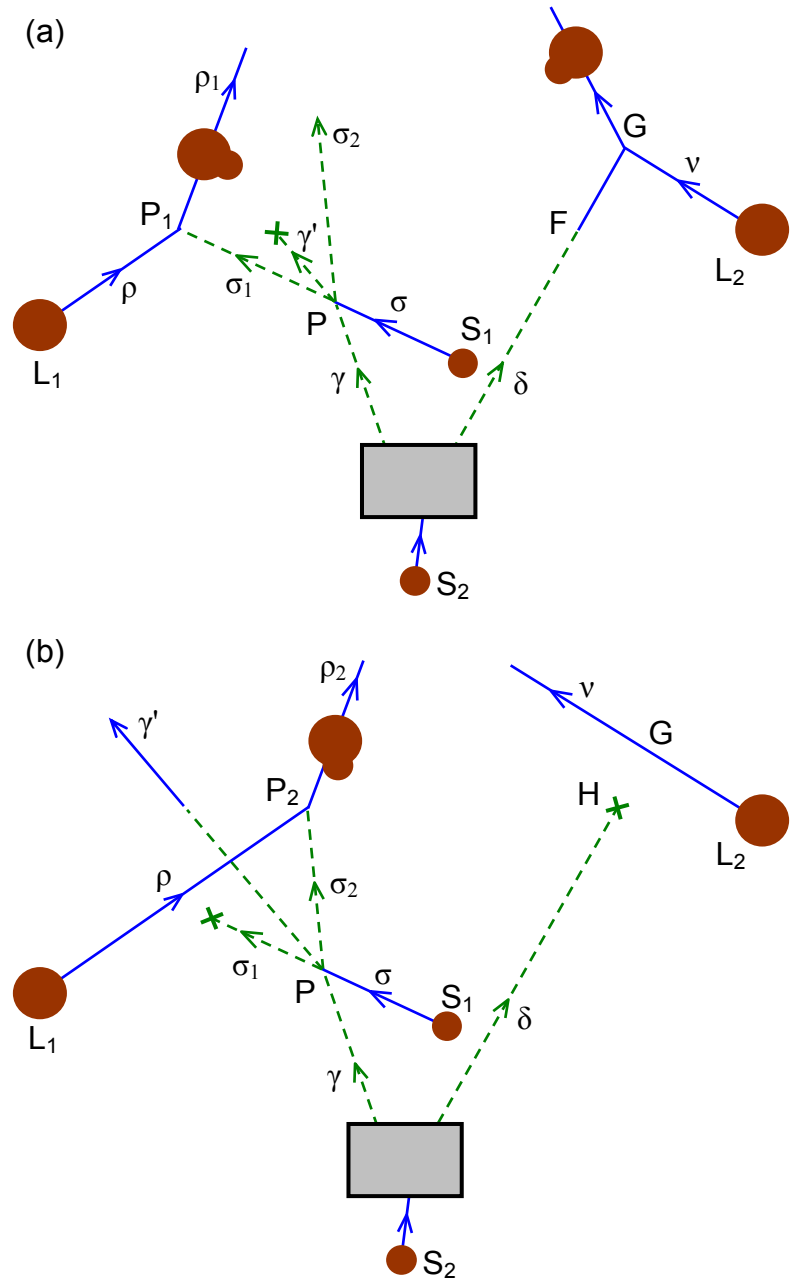

Fig. 5. (Color online) An imaginary experiment with four classical balls in the spacetime with a wormhole serves as an analog of EPR experiment in quantum mechanics. (a) and (b) Four classical balls move in the spacetime with a static wormhole: the large balls $L_1$ and $L_2$ with the unambiguous trajectories $\rho$ and $\nu$, respectively; the small ball $S_1$ with initially unambiguous trajectory $\sigma$; and the small ball $S_2$ that (after its traverse through the wormhole within grey box) moves along the ambiguous trajectories $\gamma$ and $\delta$. The initially unambiguous trajectory $\sigma$ of the small ball $S_1$ meets the ambiguous trajectory $\gamma$ of the small ball $S_2$ at point $P$ where the trajectory $\sigma$ transforms into the two ambiguous trajectories $\sigma_1$ and $\sigma_2$. After that, the system under examination has the following two variants of its further evolution. (a) According to the first variant, the small ball $S_1$ and the large ball $L_1$ collide at point $P_1$ where the ball $S_1$ is absorbed by the ball $L_1$, and then the ball $L_1$ moves along the trajectory $\rho_1$. In addition, the ball $S_2$ is absorbed by the ball $L_2$ at point $G$. For definiteness, the picture is shown in the coordinate system where the traverse of the ball $L_1$ through point $P_1$ precedes the traverse of the large ball $L_2$ through point $G$. (b) In the second evolution variant, the small ball $S_1$ and the large ball $L_1$ collide at point $P_2$ where the ball $S_1$ is absorbed by the ball $L_1$, and then the ball $L_1$ moves along the trajectory $\rho_2$. Besides, the ball $S_2$ moves along the trajectory $\gamma'$ (that becomes unambiguous at time moment at which the ball $S_1$ is absorbed by the ball $L_1$), and it is not absorbed by the ball $L_2$ at point $G$. The picture is shown in a coordinate system where the traverse of the ball $L_1$ through point $P_2$ precedes the traverse of the ball $L_2$ through point $G$.

## V. DISCUSSION. CONCLUDING REMARKS

In classical physics of spacetimes with wormhole-based time machines, there are self-inconsistent evolutions of classical balls (Fig. 2). Self-inconsistent evolutions violate the conventional causality and thereby generate a fundamental causality paradox like the grandfather paradox in science fiction. In order to avoid the causality paradox, Friedman et al [3] suggested that self-inconsistent evolutions are not allowed in nature. Deutsch [5] considered self-inconsistent evolutions of quantum systems in spacetimes with CTCs and found that they do not generate the causality paradox in the Everett multiworld interpretation of quantum mechanics. Roughly speaking, Friedman et al [3] suggest the causality paradox to exclude from reality, whereas Deutsch [5] extends reality through multiplication of the Universe in order to realize seemingly inconsistent events in different clones of the Universe.

In this paper, the alternative approach (briefly discussed earlier [11]) is developed which treats the causality paradox to be realized through behavioral uncertainties of quantum microparticles. The alternative approach is concerned with small and large classical balls whose traverse through wormholes (having finite sizes of their mouths) is permitted and forbidden, respectively. Self–inconsistent evolutions of small classical particles traversing through CTCs are specified by uncertainty. Within the suggested alternative approach, this uncertainty manifests itself in collisions of small balls with large classical balls whose evolutions are unambiguous (Fig. 3). As a result of a collision of a small ball with ambiguous trajectories and a large ball with its unambiguous trajectory, both the balls have the same unambiguous trajectory selected from its two potential versions (Fig. 3). The logical structure inherent to our interpretation of the interactions/collisions between small and large balls in spacetimes with CTCs (Fig. 3) is analogous to intrinsic logics specifying quantum measurement - wave function reduction – events in quantum mechanics outside Everett interpretation. More precisely, small and large classical balls in a spacetime with CTCs serve as analogs of microparticles and macroscopic objects in quantum mechanics, respectively. In this context, in at least a rough approximation, physics of classical balls in the spacetime with CTCs in its basis can be interpreted in terms of quantum mechanics and vice versa. In the framework of the interpretation, in terms of classical physics of solid balls moving in the spacetime with CTCs, classical analogs of quantum entanglement and EPR experiment are defined and theoretically described (Figs. 4 and 5, respectively).

Note that the interpretation is based on qualitative similarity between classical physics of solid balls moving in the spacetime with CTCs and quantum mechanics. There are many unsolved questions concerning quantitative description of self-inconsistent evolutions of classical small balls and their collisions with large balls. For instance, calculation of probabilities that characterize selection of evolution variants of small classical balls in their collisions with large balls (Fig. 3) is questionable. Also, continuous uncertainties in spatial coordinates of microparticles as well as their internal degrees of freedom are not involved in consideration in our approximate qualitative approach dealing with simplest systems of classical solid balls. A more detailed, quantitative similarity between classical physics of solid balls moving in the spacetime with CTCs and quantum mechanics will be the subject of our further research.

At the same time, the suggested qualitative interpretation logically explains the difference between microscopic and macroscopic classical balls as that related to finite sizes of wormhole mouths. (As it was noted by Penrose [20], this difference is one of key unsolved problems in fundamentals of quantum mechanics.) In this context, one can speculate that there were wormholes with microscopic mouths in the past, and their operation produced currently observed dramatic difference in behaviors of microparticles and macroscopic systems. In particular, behavioral uncertainties of quantum microparticles are treated as those resulted from effects of wormholes on small classical balls in the past. In other words, in spacetimes with wormhole-based CTCs, small classical balls – microparticles – are in their paradoxical states with no conventionally defined past, present and future.

**Acknowledgments.** This work was supported by the Ministry of Education and Science of Russian Federation (Contract 14.740.11.0353).